\documentclass[lettersize,journal]{IEEEtran}

\usepackage{amsmath,amsfonts}
\usepackage{algorithmic}
\usepackage{algorithm}

\usepackage{graphicx} % 图片插入
\usepackage{float}    % 浮动体控制

\usepackage{array}
\usepackage[caption=false,font=normalsize,labelfont=sf,textfont=sf]{subfig}
\usepackage{textcomp}
\usepackage{stfloats}
\usepackage{url}
\usepackage{verbatim}
\usepackage{graphicx}
\usepackage{enumitem}
\usepackage{amsmath}
\usepackage{booktabs}
\usepackage{multirow}
\usepackage{multicol}
\usepackage[normalem]{ulem}
\useunder{\uline}{\ul}{}
\usepackage{balance}
\usepackage{makecell}
\usepackage{xcolor}
\usepackage[marginal]{footmisc}
\definecolor{bgcolor}{RGB}{242, 242, 242}
\usepackage{caption}
\usepackage{marvosym}

\begin{document}

\title{Cloud-Device Collaborative Agents for \\ Sequential Recommendation}

\author{
    \IEEEauthorblockN{Jing Long, Sirui Huang, Huan Huo, Tong Chen, Hongzhi Yin, Guandong Xu$^{*}$}
\thanks{Jing Long, Sirui Huang, and Huan Huo are affiliated with the Data Science and Machine Intelligence Lab, University of Technology Sydney, Sydney, Australia. Tong Chen, and Hongzhi Yin are with the School of Electrical Engineering and Computer Science, the University of Queensland, Brisbane, Australia. Guandong Xu is with the Facility of Data Science and Artificial Intelligence, The Education University of Hong Kong, Hong Kong, China.
}
\thanks{Corresponding author: Guandong Xu$^{*}$ (email: gdxu@eduhk.hk)}

}





\maketitle

% Done
\begin{abstract}
Recent advances in large language models (LLMs) have enabled agent-based recommendation systems with strong semantic understanding and flexible reasoning capabilities. While LLM-based agents deployed in the cloud offer powerful personalization, they often suffer from privacy concerns, limited access to real-time signals, and scalability bottlenecks. Conversely, on-device agents ensure privacy and responsiveness but lack the computational power for global modeling and large-scale retrieval. To bridge these complementary limitations, we propose CDA4Rec, a novel Cloud-Device collaborative framework for sequential Recommendation, powered by dual agents: a cloud-side LLM and a device-side small language model (SLM). CDA4Rec tackles the core challenge of cloud-device coordination by decomposing the recommendation task into modular sub-tasks including semantic modeling, candidate retrieval, structured user modeling, and final ranking, which are allocated to cloud or device based on computational demands and privacy sensitivity. A strategy planning mechanism leverages the cloud agent’s reasoning ability to generate personalized execution plans, enabling context-aware task assignment and partial parallel execution across agents. This design ensures real-time responsiveness, improved efficiency, and fine-grained personalization, even under diverse user states and behavioral sparsity. Extensive experiments across multiple real-world datasets demonstrate that CDA4Rec consistently outperforms competitive baselines in both accuracy and efficiency, validating its effectiveness in heterogeneous and resource-constrained environments.
\end{abstract}

\begin{IEEEkeywords}
Recommender Systems, Cloud-Device Collaboration, Multi Agents, Large Language Models
\end{IEEEkeywords}

\section{Introduction}

% Done
In the modern era, sequential recommender systems have become essential tools for assisting users in discovering relevant content and making informed decisions across various domains. Traditional approaches, such as those based on recurrent neural networks (RNNs) \cite{xu2021long,han2021deeprec,long2023model} and graph neural networks (GNNs) \cite{he2020lightgcn,gao2022graph}, have demonstrated effectiveness in modeling behavioral patterns from historical interactions. However, these models often rely on rigid architectures and have difficulty in handling diverse user intents or rich contextual signals. To address these limitations, recent studies have introduced Large Language Models (LLMs) into sequential recommendation \cite{lin2024data,zhao2024llm}, leveraging their semantic understanding and prompt-based reasoning to offer more flexible and generalizable solutions. Building on this trend, recent studies have proposed using LLM-based agents for recommendation \cite{shu2023rah,wang2024macrec}, where the LLM serves not just as a generator but as an autonomous decision-maker capable of reasoning over user intent, selecting relevant items, and planning recommendation strategies. This agent formulation enables more flexible and adaptive recommendation pipelines, allowing the system to dynamically incorporate user context and generate interpretable outputs. In this way, LLM-based agents offer stronger adaptability and decision-making capabilities than standard LLM models, enabling more personalized, context-aware, and goal-directed recommendations.

\begin{figure}
    \centering  
	\includegraphics[width=\linewidth]{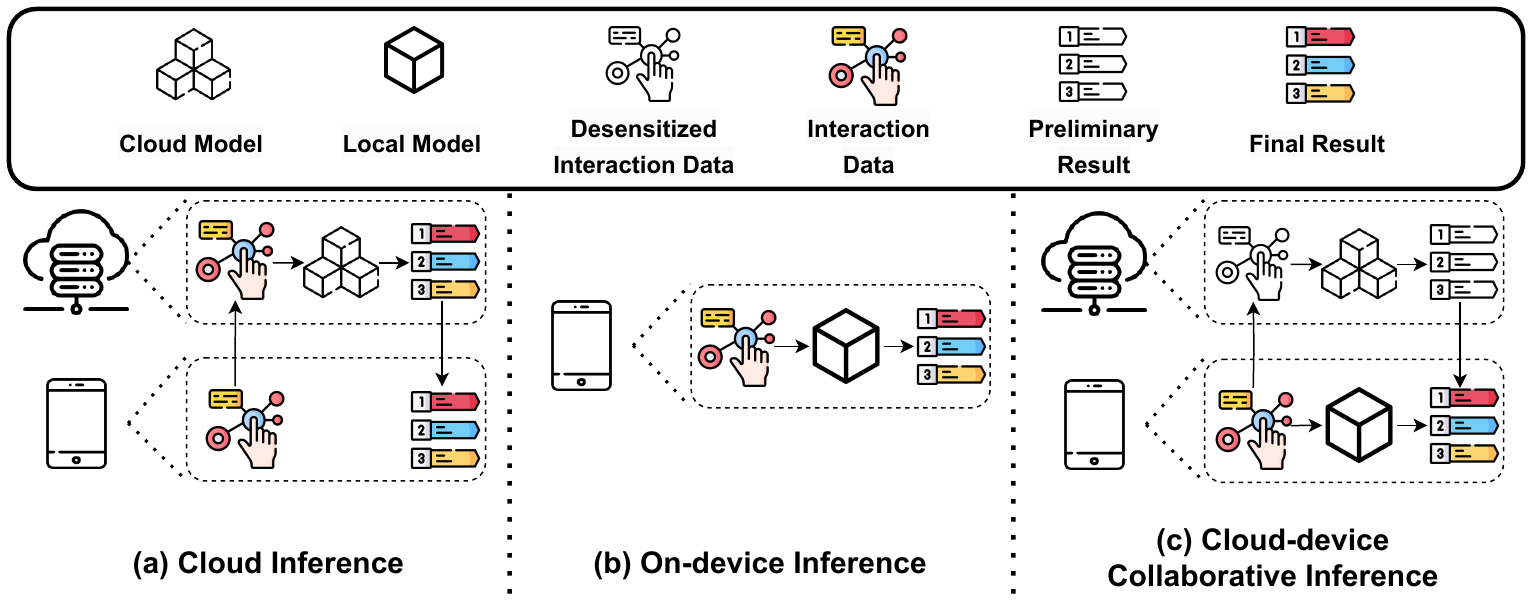}
	\caption{(a) Example of Cloud Inference. (b) Example of on-device Inference. (c) Example of Cloud-device Collaborative Inference.}
	\label{Fig:toyexample} 
\end{figure}

While LLM-based agents have shown promising capabilities in personalized recommendation, most existing implementations are deployed in the cloud as illustrated in Figure \ref{Fig:toyexample} (a). In this setting, user requests and behavioral data are uploaded to centralized servers, where large models process the information and return recommendation results. Despite their effectiveness, these systems face several limitations. Due to privacy concerns and transmission constraints, they cannot directly access real-time user signals, which hampers accurate modeling of fine-grained and dynamic user intent \cite{han2021deeprec,xia2022device}. In addition, as the number of users grows, they often experience computational bottlenecks, making it difficult to allocate sufficient resources for timely and fully personalized inference at scale. To address these limitations, recent research has explored deploying Small Language Models (SLMs) directly on user devices \cite{xu2024device}, enabling on-device agents to perform local inference, as shown in Figure \ref{Fig:toyexample}(b). This design allows direct access to user data for efficient personalization and alleviates server-side computation load. Unfortunately, these methods are inherently constrained by the limited computational and memory resources of user devices, which hinders their ability to perform large-scale retrieval with satisfactory accuracy. Consequently, the subpar recommendation performance largely offsets the convenience and efficiency of purely on-device LLM agents.

Given the complementary yet limited capabilities of cloud-based and on-device recommendation agents, it is essential to explore new frameworks that can integrate their advantages while mitigating their respective drawbacks. Preliminary efforts have explored the potential of edge-cloud collaboration, where a cloud-side agent and a device-side agent work cooperatively to complete complex natural language processing tasks \cite{shao2025division}. Inspired by this direction, we propose CDA4Rec, a novel \textbf{C}loud-\textbf{D}evice collaborative framework that leverages LLM-empowered \textbf{A}gents for sequential \textbf{Rec}ommendation. As illustrated in Figure~\ref{Fig:toyexample}(c), CDA4Rec employs a dual-agent architecture, where the cloud agent handles computationally intensive and less privacy-sensitive tasks, while the device agent focuses on lightweight computations involving real-time and more sensitive user data. This collaborative setup aims to combine the strengths of both sides by leveraging the cloud's capacity for global reasoning and the device’s access to immediate contextual signals, while mitigating their respective limitations.

Despite this, it remains highly challenging to instantiate such a collaborative recommendation framework in practice. A key difficulty lies in determining \textbf{how the cloud and device agents should cooperate to complete the recommendation task}. Specifically, the system must determine how to decompose the overall recommendation task into meaningful sub-tasks, and further decide which of these should be executed by the cloud agent and which should be delegated to the device agent. These decisions must also consider user privacy constraints, as it is impractical to assume the cloud's access to sensitive behavioral signals and real-time contextual data generated on-device.

This challenge becomes even more pronounced when \textbf{user heterogeneity} is taken into account. That is, each user exhibits distinct recommendation needs. For example, active users with rich and consistent interaction histories can benefit from fine-grained personalization that leverages long-term preferences and behavioral patterns. In contrast, inactive users or cold-start users lack sufficient behavioral signals for accurate modeling and therefore require more exploratory or diversity-promoting strategies to infer their interests. Additionally, users may differ in how clearly they express their intent. Some demonstrate strong, goal-directed behavior, while others engage in more ambiguous or exploratory interactions. These variations further complicate the realization of cloud-device collaborative recommendation, as the system must operate under constraints of fragmented data access and imbalanced agent capabilities. Different from cloud-based agents which can centrally observe and process most user information, cloud-device collaborative frameworks must reason over incomplete and distributed user signals. This fragmentation makes it substantially more difficult to maintain consistent recommendation quality across diverse user states. To summarize, the core challenge lies in determining how to effectively divide and coordinate recommendation tasks between cloud and device agents, while accounting for privacy constraints and the diverse states and needs of users.

To address this challenge, CDA4Rec adopts a modular execution flow guided by a strategy planning mechanism, enabling fine-grained coordination between cloud and device agents. The overall recommendation is decomposed into a series of sub-tasks including semantic modeling, candidate retrieval, structured user modeling, and final ranking. Among them, semantic modeling and candidate retrieval involve high computational cost and rely on globally available knowledge with minimal privacy concerns, and are thus assigned to the cloud agent. In contrast, structured modeling and final ranking require access to sensitive behavioral signals and real-time contextual data, which are only available on the device, making them more suitable for the device agent. Importantly, semantic modeling on the cloud and structured modeling on the device can be executed in parallel, significantly reducing end-to-end latency.

To handle the diverse needs introduced by user heterogeneity and to support effective cloud-device collaboration, we additionally introduce a strategy planning mechanism that enables the cloud agent to dynamically generate a personalized execution plan based on a privacy-preserving user abstract provided by the device. This plan determines which sub-tasks to activate and how to execute them, depending on factors such as query specificity, behavioral sparsity, and domain context. Built upon this framework, CDA4Rec enables efficient and adaptive collaboration between agents, effectively balancing recommendation accuracy, system latency, and privacy through task-aware planning and parallel execution. Our main contributions are summarized as follows:
\begin{itemize}[leftmargin=*]
    \item We propose CDA4Rec, a novel cloud-device collaborative recommendation framework built upon a dual-agent architecture. It empowers a cloud-based LLM and an on-device SLM to collaboratively perform personalized recommendations through modular execution and agent-driven coordination. The framework supports partial parallel execution across agents to improve efficiency.

    \item We design a strategy planning mechanism that enables the cloud agent to dynamically generate personalized execution plans. These plans determine which sub-tasks to activate and how to assign them across cloud and device, considering factors such as query specificity, behavioral sparsity, and privacy sensitivity.

    \item We conduct extensive experiments on multiple real-world datasets across different domains. Results show that CDA4Rec consistently outperforms strong baselines in both recommendation accuracy and computational efficiency, demonstrating its robustness under diverse user states.
\end{itemize}

\begin{figure*}
	\includegraphics[width=\linewidth]{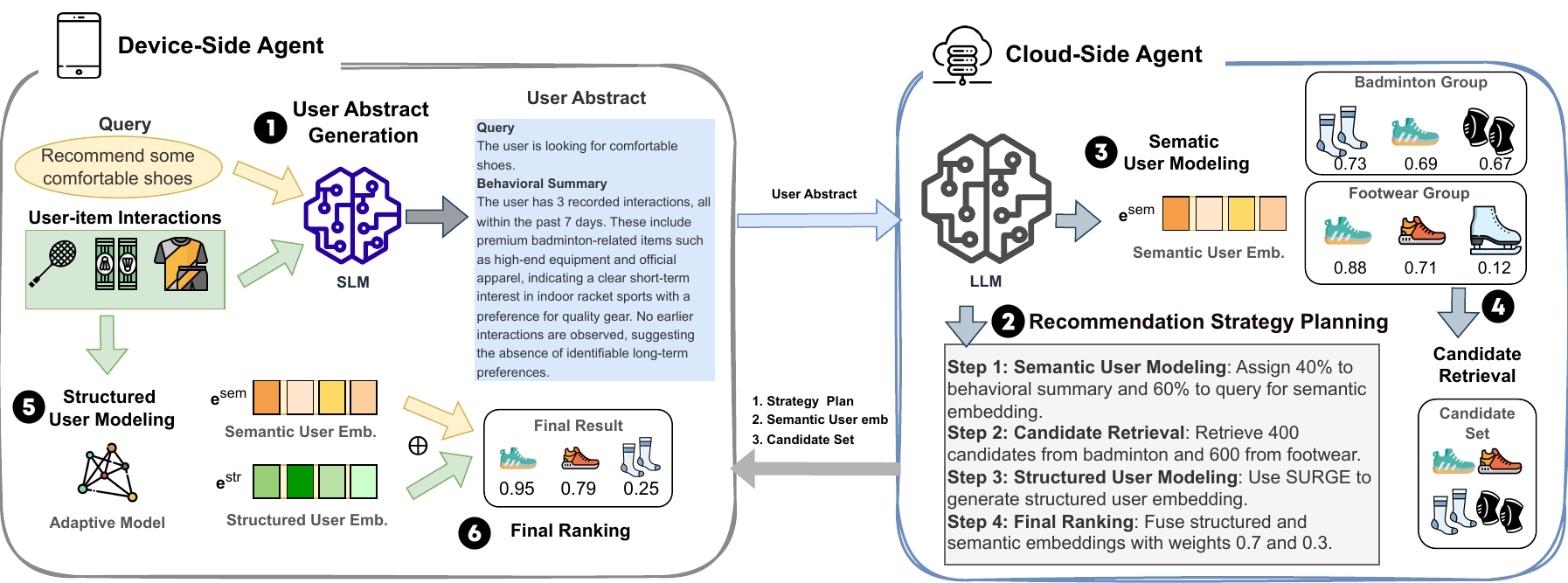}
    \caption{The overview of our proposed CDA4Rec.}
	\label{fig:overview} %% label for entire figure 
\end{figure*}

\section{Preliminaries}
% Done
\noindent\textbf{User.} 
Each user $u \in \mathcal{U}$ is associated with a recommendation query $q_u$ and a historical interaction sequence $\mathcal{X}_u$. The query $q_u$ is a natural language request that reflects the user’s current intent. The interaction history $\mathcal{X}_u = \{(i_1, t_1), (i_2, t_2), \ldots, (i_M, t_M)\}$ denotes a time-ordered sequence of item interactions, where $i_m \in \mathcal{I}$ and $t_m$ are the item ID and the timestamp of the $m$-th interaction.

% Done
% \noindent\textbf{Item.} 
% Each item $i \in \mathcal{I}$ is associated with an embedding vector $\textbf{e}_i \in \mathbb{R}^d$, which can be initialized by applying the cloud-based large language model $\Theta_l$ to analyze the item's textual metadata $\text{text}(i)$ (e.g., title, brand, tags):
% \begin{equation}
%     \textbf{e}_i = \text{AVG}(\Theta_l(text(i))),
% \end{equation}
% where $\text{AVG}(\cdot)$ denotes mean pooling over the final-layer token representations. On this basis, we use $\textbf{E}_{\mathcal{I}}$ to denote the embedding matrix for all items. 
% change the embedding size

\noindent\textbf{Item.} 
Each item $i \in \mathcal{I}$ is associated with an embedding vector $\textbf{e}_i \in \mathbb{R}^d$, which is initialized by applying the cloud-based large language model $\Theta_l$ to analyze the item's textual metadata $\text{text}(i)$ (e.g., title, brand, tags):
\begin{equation}
    \textbf{e}_i = \text{PCA}(\text{AVG}(\Theta_l(\text{text}(i)))).
\end{equation}
Here, $\text{PCA}(\cdot)$ denotes a dimensionality reduction step applied offline, which compresses the original high-dimensional LLM output into the target embedding space $\mathbb{R}^d$. The resulting embeddings are stored as the fixed item representation matrix $\textbf{E}_{\mathcal{I}}$ for downstream training.

% Done
\noindent\textbf{Sequential Recommendation.} 
Given a recommendation query $q_u$ and a historical interaction sequence $\mathcal{X}_u$, our goal is to recommend a ranked list of items that best matches the user's current intent efficiently. In addition, the system can optionally generate natural language explanations to clarify the rationale behind each suggestion.

\section{Methodology}

\subsection{Framework Overview}

% Done
CDA4Rec is a cloud-device collaborative recommendation framework built upon a dual-agent architecture. Instead of following a static recommendation pipeline, CDA4Rec empowers two intelligent agents: (1) a \textbf{device-side agent} powered by a lightweight small language model (SLM), and (2) a \textbf{cloud-side agent} powered by a powerful large language model (LLM). These agents collaboratively plan and execute personalized recommendations in a modular and adaptive fashion. Each agent is capable of perceiving context, making decisions, and selectively activating modules based on task requirements, which enables CDA4Rec to dynamically adapt to diverse recommendation scenarios.

% Done
As shown in Figure \ref{fig:overview}, the recommendation process begins when a user submits a query. The device-side agent first analyzes the query and local interaction history to generate a privacy-respecting user abstract, composed of a sanitized query and a behavioral summary. This abstract is then transmitted to the cloud-side agent, which serves as a strategy planner. Leveraging its reasoning abilities, the LLM analyzes the abstract and domain context to formulate a personalized recommendation plan, which specifies which modeling modules to activate and how to execute tasks across cloud and device. Following the strategy plan, the cloud agent performs semantic user modeling and coarse candidate retrieval based on the abstracted intent. In parallel, the device-side agent may execute structured user modeling using local interaction history. Once both representations are ready, the device completes the final ranking by integrating structured and semantic signals. Optionally, the device agent can also generate natural language explanations for the recommendation results to enhance transparency and user trust.

% Done
By orchestrating these modules in an agent-driven and context-aware manner, CDA4Rec offers a flexible and interpretable recommendation process while significantly improving computational efficiency through concurrent cloud-device collaboration.

\subsection{Abstract Generation}
% Done
To facilitate semantic reasoning and personalized strategy planning on the cloud while maintaining user privacy, CDA4Rec assigns the responsibility of contextual summarization to the device-side agent. Rather than transmitting raw queries or detailed interaction logs, the agent proactively interprets the user’s current state and condenses it into a structured natural language abstract. This abstract acts as a privacy-aware communication interface, allowing the cloud-side agent to operate with sufficient behavioral context without directly accessing sensitive information.

% What is the input to the on-device SLM for summary generation? All historical interactions? Any side information used or only the item titles?

\begin{figure}
	\includegraphics[width=\linewidth]{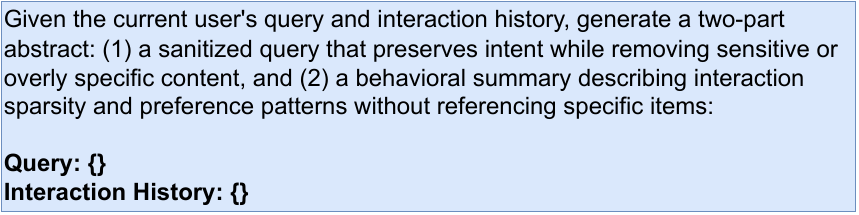}
    \caption{The prompt for abstract generation}
	\label{fig:prompt_ag} %% label for entire figure 
\end{figure}

% Done
Upon receiving a recommendation query $q_u$, the device-side agent powered by the SLM $\Theta_s$ takes initiative to generate a personalized abstract $a_u$ that encapsulates key user signals in a concise form. This abstract consists of two parts including a sanitized query $\hat{q}_u$ and a behavioral summary $b_u$. The sanitized query is a rewritten version of $q_u$ that preserves its semantic intent while removing identifiers or overly specific personal content. The behavioral summary reflects high-level user dynamics derived from $\mathcal{X}_u$, including interaction sparsity, long-term preference trends, and recent short-term interests, without referencing specific items. This reasoning process is executed via a prompt-driven interface, where the agent uses a structured instruction prompt shown in Figure \ref{fig:prompt_ag} to guide the generation process, ensuring modularity and consistency in agent behavior.

% Done
Notably, although each interaction in $\mathcal{X}_u$ is stored as an item ID with the interaction timestamp in the system, we retrieve the corresponding item descriptions (e.g., title, brand, tags) in advance and include them directly in the prompt, allowing the agent to access rich semantic information. The constructed abstract $a_u$ is then transmitted to the cloud-side agent, serving as the basis for downstream planning and modeling.

\subsection{Strategy Planning}\label{sec:sp}

% Done
In CDA4Rec, the recommendation process is coordinated by a cloud-side agent powered by a large language model (LLM), which assumes responsibility for global planning. Rather than executing a fixed pipeline, this agent actively reasons about the user’s recommendation context and produces an adaptive strategy tailored to the user’s needs.

% Done
Upon receiving the personalized user abstract $a_u$ from the device-side agent, the cloud agent analyzes key contextual signals (i.e., query specificity, interaction sparsity, and domain characteristics) to assess the complexity and intent behind the recommendation task. Based on this assessment, it generates a structured recommendation plan $\pi_u$ that determines how each downstream module should be executed, and under what configuration. The strategy plan includes not only execution parameters, but also explicit justifications that explain why each decision is made. These justifications enhance transparency and facilitate consistent execution across agents.

% Done
Apart from abstract generation and strategy planning, CDA4Rec defines four core modules for completing a recommendation task. While all planning decisions are made by the cloud-side agent, execution is distributed between the cloud and device agents, ensuring both personalization and efficiency. In what follows, we first discuss the planning logic of each module, and then integrate these factors into a structured prompt, enabling the LLM to generate interpretable and reliable execution plans:

\begin{itemize}[leftmargin=*]

% Fig 4 contains some notations that are not introduced in this part, e.g., N for tags and \beta. Their meanings should be made clear before entering subsequent sections.
% Done
\item \textbf{Semantic User Modeling.}
This module constructs a semantic user embedding to support intent-aware candidate retrieval and final ranking. It takes as input the user abstract, which includes a sanitized query and a behavioral summary. The relative informativeness of these two components is critical for generating a high-quality embedding. Specifically, a clearly stated query should be assigned greater weight, while a rich and recent interaction history warrants more emphasis on the behavioral summary. To this end, the cloud-side agent first evaluates these signals during the planning stage and determines their relative importance accordingly. The output is a weighting coefficient $\alpha$, where $\alpha$ and $1 - \alpha$ represent the importance of the query and the behavioral summary, respectively.

% Done
\item \textbf{Candidate Retrieval.}
This module generates a relevant candidate set from the large-scale item corpus to support efficient ranking on the device side. Computing similarities between the semantic embedding across all items is prohibitively expensive, and thus, CDA4Rec adopts a two-step approach. It first leverages intent and behavioral cues to identify a set of tag groups that are relevant to the user, and then retrieves items exclusively from these selected groups. However, a uniform retrieval settings for all users, such as a fixed number of tags or candidates, may fail to reflect user-specific intent or waste resources on excessive or irrelevant items. 

To address this, the cloud-side agent dynamically configures the retrieval strategy considering query clarity, behavior summary and domain characteristics, which can significantly affects the candidate level. For example, when the query clearly conveys the user’s intent and recent behaviors are rich, a smaller, more focused retrieval scope may suffice. Conversely, in cases of ambiguous intent or sparse history, the agent broadens retrieval to ensure sufficient coverage. Domain characteristics also influence decisions where dynamic domains like news or music favor larger, more diverse candidate sets, while static domains such as movies benefit from compact, topic-focused pools. To ensure consistency and interpretability, the cloud-based agent selects from predefined tag counts \{3, 5, 10, 20\} and candidate sizes \{500, 1000, 2000, 5000\}, and further assigns retrieval weights to each tag group to guide proportional item sampling.

The final retrieval configuration can be represented as a weighted tag dictionary $\mathcal{N}_u = \{tag_1: N_1, tag_2: N_2, \ldots\}$, where $tag_i$ denotes a selected tag group and $N_i$ specifies the number of items to retrieve from that group.

% Done
\item \textbf{Structured User Modeling.}
This module generates a structured user embedding based on the user's sequential interaction history, which captures fine-grained behavioral patterns and enhances the final ranking of candidate items. After proceeding, the cloud-side agent first decides whether structured modeling is necessary by inspecting the user's interaction density and recency. If modeling is warranted, the cloud-side agent selects an encoder from SASRec \cite{kang2018self} or SURGE \cite{chang2021sequential}. SASRec excels at capturing clear sequential patterns via self-attention, while SURGE models complex, non-linear item transitions through graphs. The parameters of the selected encoder are optimized using the training objective defined in Equation \ref{eq10}.

% It should be made clearer that this part uses results from Candidate Retrieval sent from the cloud, not the entire item set. This will enhance your efficiency statement.
% Done
% \item \textbf{Final Ranking.}
% This module fuses the semantic and structured user embeddings to rank the retrieved candidate items. To determine the optimal fusion strategy, the cloud-side agent assesses the reliability of each embedding based on query clarity and interaction recency. For instance, when the query is specific but past behavior is sparse, the agent assigns more weight to the semantic embedding; conversely, rich and recent behavior leads to a behavior-driven fusion. 
% \end{itemize}

\item \textbf{Final Ranking.}
This module ranks the candidate items retrieved from the cloud-side agent, rather than scoring the entire item corpus, which significantly improves inference efficiency. It fuses the semantic and structured user embeddings to compute relevance scores for each candidate. To determine the optimal fusion strategy, the cloud-side agent evaluates the reliability of each embedding based on query clarity and interaction recency. Specifically, it outputs a weighting coefficient $\beta$, where a higher $\beta$ indicates greater emphasis on the semantic embedding derived from query and behavioral summaries, and $1 - \beta$ reflects the weight assigned to the structured embedding constructed from sequential user behaviors. This fused representation is then used to rank the candidate items on the device side.
\end{itemize}

\begin{figure}
	\includegraphics[width=\linewidth]{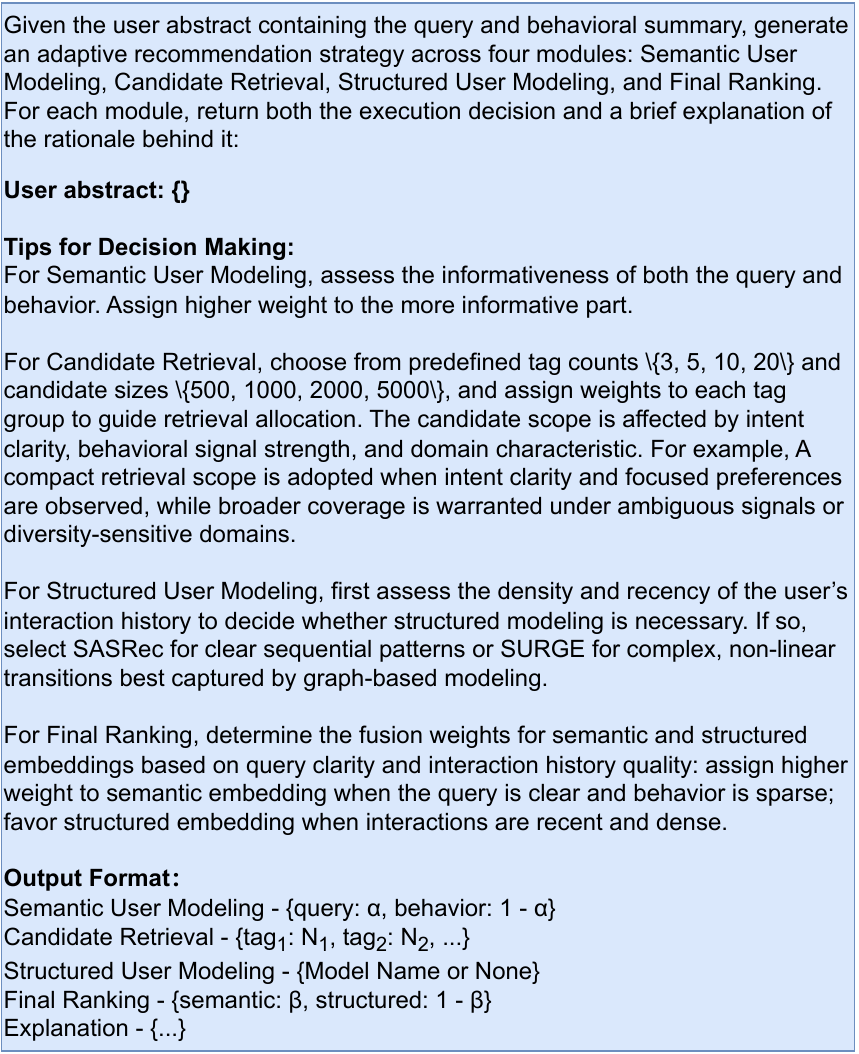}
    \caption{The prompt for strategy planning}
	\label{fig:prompt_sp} %% label for entire figure 
\end{figure}

% Done
% To realize this reasoning process, CDA4Rec employs a prompt-driven interface that instructs the cloud-side agent to analyze the the abstract and return a structured plan. The prompt shown in Figure \ref{fig:prompt_sp} includes a breakdown of each module's decision criterial and enforces a unified output format for ease of paring and downstream execution.
To realize this reasoning process, CDA4Rec employs a prompt-driven interface that instructs the cloud-side agent to analyze the user abstract and return a structured plan. The prompt, shown in Figure \ref{fig:prompt_sp}, includes a breakdown of each module’s decision criteria and enforces a unified output format for ease of parsing and downstream execution. The final output of this step is the personalized strategy plan $\pi_u$, which specifies module activation, and key parameters for subsequent recommendation stages.

% It's not clear how user heterogeneity is handled here - is it done by learning this alpha value per user? In which case more weight is given to the query, and in which case more weight is given to the behavioral information? This should be emphasized. 
% Done
\subsection{Semantic User Modeling}
After generating the personalized plan $\pi_u$, the cloud-side agent then utilizes the LLM $\Theta_l$ to derive the semantic user embedding from the uploaded user abstract $a_u$ which contains the sanitized query $q_u$ versus the behavioral summary $b_u$:
\begin{equation}
    \textbf{e}_u^{sem} = \alpha \cdot AVG(\Theta_l(\hat{q}_u)) + (1-\alpha) \cdot AVG(\Theta_l(b_u)),
\end{equation}
where $\mathrm{AVG}(\cdot)$ denotes mean pooling over the final-layer token representations, and $\alpha \in [0,1]$ is obtained from the personalized plan $\pi_u$ which specifies the relative importance of $q_u$ and $b_u$.

% Done
\subsection{Candidate Retrieval}
Given the semantic user embedding $\textbf{e}_u^{sem}$ from the previous stage and candidate retrieve plan $\mathcal{N}_u = \{tag1:N_1, tag2:N_2,...\}$, the cloud-side agent retrieve a customized candidate set from the large-scale item corpus. For each tag, we compute the similarity between the semantic user embedding with each item embedding for all items within this tag:
\begin{equation}
\hat{y}_i = \textbf{e}^{sem}_u \cdot \textbf{e}_i.
\end{equation}
Based on the similarity scores, we select the top-$N$ items from each tag group. The final candidate $\mathcal{C}_u$ set is obtained by aggregating the top-scoring items across all selected tag groups.
% how about the loss

% Done
\subsection{Structured User Modeling}
To capture personalized behavioral preferences, the device-side agent derives a structured user embedding $\textbf{e}_u^{str}$ from the user's domain-specific interaction history $\mathcal{X}_u = \{(i_1, t_1), (i_2, t_2), \ldots, (i_M, t_M)\}$:
\begin{equation}
    \textbf{e}_u^{str} = f_{loc}(\mathcal{X}_u),
\end{equation}
where $f_{loc}(\cdot) \in\{\text{SASRec}, \text{SURGE}\}$ refers to the behavioral modeling function, indicated by the personalized plan. 

% Since the computation relies on item embeddings stored in the cloud, the device must first query these embeddings. To preserve user privacy, the request is obfuscated by mixing in a set of unrelated item IDs, ensuring the cloud cannot infer the user's true interaction history. 
% Not sure about offline recommendation

% Done
\subsection{Final Ranking}
Given the user's semantic embedding $\textbf{e}_u^{sem}$ and structured embedding $\textbf{e}_u^{str}$, the device-side agent fuses these two representations to capture both intent-aware semantics and historical behavioral signals under the guidance of the personalized plan $\pi_u$. Specifically, the fused user embedding is obtained by:
\begin{equation}
\textbf{e}_u = \beta \cdot \textbf{e}_u^{sem} + (1 - \beta) \cdot \textbf{e}_u^{str},
\end{equation}
where $\beta \in [0, 1]$ is exported from the personalized plan $\pi_u$ which controls the relative importance of semantic and behavioral signals. The fused user embedding is then compared with each item embedding $\textbf{e}_i$ in the candidate set $\mathcal{C}_u$ to compute final relevance scores:
\begin{equation}
\hat{y}_i = \textbf{e}_u \cdot \textbf{e}_i.
\end{equation}
The top-$K$ items based on the ranking scores form the final recommendation list:
\begin{equation}
\mathcal{R}_u = \text{TopK}\left( \{ \hat{y}_i \mid i \in \mathcal{C}_u \} \right).
\end{equation}

% Done
\subsection{Explanation Generation}
\begin{figure}
	\includegraphics[width=\linewidth]{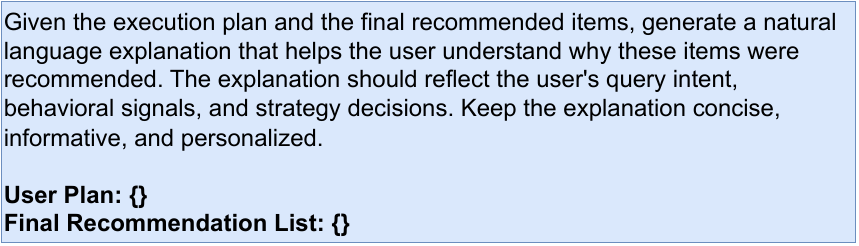}
    \caption{The prompt for explanation generation}
	\label{fig:prompt_eg} 
\end{figure}

To enhance transparency and foster user trust, the device-side agent supports the optional generation of natural language explanations $\mathcal{E}_u$ for the final recommendation results, with the choice to enable this feature left entirely to the user. The generation process is implemented by the lightweight on-device SLM $\Theta_s$ with a structured prompt as shown in Figure \ref{fig:prompt_eg}, which is constructed using the personalized plan with execution logs $\pi_u$, and final recommendation list with related item descriptions $\mathcal{R}_u$.

\subsection{Model Optimization}
% Done
Optimizing CDA4Rec focuses on fine-tuning the lightweight SLM and training item embeddings to align device-side behaviors with cloud-side reasoning. This alignment enables effective abstract generation, interpretable recommendation, and accurate item matching. Notably, the cloud-based LLM is not updated during training, as it serves as a powerful, general-purpose model whose reasoning capabilities are utilized in a zero-shot manner to ensure stability and reduce deployment costs. All training is performed on the cloud using public or non-sensitive data, after which the optimized components are deployed to the device. This design preserves user privacy by keeping sensitive data local, while benefiting from centralized training efficiency and strong generalization.

% Done
At first, CDA4Rec introduces two trainable QLoRA \cite{dettmers2023qlora} adapters $\phi_{\text{abs}}$ and $\phi_{\text{exp}}$ for abstract generation and explanation generation respectively. The \textbf{abstract generation adapter} $\phi_{\text{abs}}$ is designed to ensure that the abstract produced by SLM leads the LLM to produce a downstream strategy plan that closely matches the one derived from full access to raw inputs. To achieve this, we simulate both processes and minimize their divergence through a cross-entropy loss:
\begin{equation}
    \mathcal{L}_{\text{abs}} = l_{\text{CE}}(\pi_u, \pi'_u),
\end{equation}
where $\pi_u = \Theta_l(a_u)$ is the plan generated by the LLM given the abstract $a_u = \Theta_s \oplus \phi_{\text{abs}}(q_u, \mathcal{X}_u)$, and $\pi'_u = \Theta_l(q_u, \mathcal{X}_u)$ is the gold-standard plan generated by the LLM using the full query and history. The \textbf{explanation generation adapter} $\phi_{\text{exp}}$ enhances the SLM's ability to produce natural language explanations for final recommendations. It is optimized by aligning the SLM’s explanation output with that of the cloud LLM using the same prompt inputs:
\begin{equation}
    \mathcal{L}_{\text{exp}} = l_{\text{CE}}(\mathcal{E}_u, \mathcal{E}'_u),
\end{equation}
where $\mathcal{E}_u = \Theta_s \oplus \phi_{\text{exp}}(\pi_u, \mathcal{R}_u)$ and $\mathcal{E}'_u = \Theta_l(\pi_u, \mathcal{R}_u)$ are the explanations generated by the SLM and LLM respectively, based on the execution plan $\pi_u$ and the final recommendation list $\mathcal{R}_u$. It is worth noting that the LLM only generates explanations during the training phase.

%It should be mentioned that the LLM only generates explanations during training, not inference.

% Done
In addition, CDA4Rec learns item embeddings and parameters from local functions (i.e., SASRec and SURGE) to support accurate item matching during candidate retrieval and final ranking. The learning objective is to align the fused user representation $\textbf{e}_u$ with relevant item representations. Specifically, we treat each interacted item in the user's history as the positive label and randomly sample negative items from non-interacted items, to formulate a contrastive training loss:
\begin{equation}\label{eq10}
\mathcal{L}_{\text{REC}} = \sum_{j\in\mathcal{N}^-}log\sigma(y_i - y_j),
\end{equation}
where $\sigma(\cdot)$ is the sigmoid function, while $y_i$ and $y_j$ are respectively the scores assigned to the positive and negative samples.

% \begin{algorithm}
%   \caption{The optimization of CDA4Rec.}
%   \label{alg:opt}
% \begin{algorithmic}[1]
%     \STATE Obtain the large language model $\Theta_l$;
%     \STATE Obtain the small language model $\Theta_s$;
%     \STATE Initialize the item embedding matrix $\textbf{E}_{\mathcal{I}}$
%     \STATE Initialize adapters $\phi_{abs}$ and $\phi_{exp}$ for abstract generation and explainable recommendation respectively;
%     \REPEAT
%         \FOR{$(q, \mathcal{X},y)\in\mathcal{D}$}
%             \STATE $a' = \Theta_s\oplus\phi_{abs}(q,\mathcal{X})$;
%             \STATE $a = \Theta_l(q,\mathcal{X})$;
%             \STATE Take a gradient step w.r.t $\mathcal{L}_{abs}(a',a)$ to update $\phi_{abs}$;
%             \STATE $\alpha, \beta, \pi = \Theta_l(a)$
%             \STATE $\textbf{e}^{sem} = \alpha \cdot AVG(\Theta_l(q_u)) + (1-\alpha) \cdot AVG(\Theta_l(b_u))$;
%             \STATE $\textbf{e}^{str} = f_{loc}(\mathcal{X})$;
%             \STATE $\textbf{e}_u = \lambda \cdot \textbf{e}_u^{\text{sem}} + (1 - \lambda) \cdot \textbf{e}_u^{\text{str}}$;
%             \STATE $y' = \textbf{E}_\mathcal{I}(\textbf{e})$;
%             \STATE Take a gradient step w.r.t $\mathcal{L}_{rec}(y',y)$ to update $\textbf{E}_\mathcal{I}$;
%             \STATE $\Lambda' = \Theta_s\oplus\phi_{exp}(\pi,\mathcal{R})$;
%             \STATE $\Lambda = \Theta_l(\pi,\mathcal{R})$;
%             \STATE Take a gradient step w.r.t $\mathcal{L}_{exp}(\Lambda',\Lambda)$ to update $\phi_{exp}$;
%         \ENDFOR
%     \UNTIL{convergence}
% \end{algorithmic}
% \end{algorithm}

\section{Experiments}
% Done
In this section, we conduct experiments to verify the effectiveness and efficiency of CDA4Rec in sequential recommendation. Specifically, we aim to answer the following research questions (RQs):

{\small
\noindent{\textbf{RQ1}}: Does CDA4Rec outperform existing baselines?\\
\noindent{\textbf{RQ2}}: How effective are the key components of CDA4Rec?\\
\noindent{\textbf{RQ3}}: How does CDA4Rec perform under different history lengths?\\
\noindent{\textbf{RQ4}}: Can CDA4Rec support few-shot recommendation for new users?\\
\noindent{\textbf{RQ5}}: Can CDA4Rec generate interpretable and user-aligned results?
}

\subsection{Dataset and Evaluation Protocols}
We evaluated our method on four widely used datasets, which cover various product domain including Movie \cite{li2018conversational}, Music \cite{santana2020music4all}, Sport\cite{ni2019justifying}, and POI \cite{yang2019revisiting}. Table \ref{tab:ds} summarizes the statistics of the two datasets. Among them, the Redial dataset provides explicit user queries, while the remaining datasets only contain user-item interaction histories. The diversity in data availability enables us to validate the robustness of our proposed framework across different recommendation scenarios. More notably, we do not consider cross-domain recommendation in this work and each user's interaction history contains only items from a single domain. We assume that the recommendation domain is known at the time of the user request, either inferred from the query itself or specified by the application context. Additionally, in our cloud-device setting, only item embeddings from relevant domains are selectively cached on the device, allowing efficient local access during inference. This approach reduces transmission latency and enhances responsiveness, while avoiding the burden of storing the full embedding matrix on resource-constrained devices.

% 更符合现实场景
For each dataset, we randomly select 10\% of users and their entire interaction histories to be excluded from training, in order to evaluate the model’s few-shot recommendation capability (RQ4). The remaining 90\% of users are used for standard training, validation, and testing. For evaluation, we use each user's last interaction as the test instance, the second-to-last for validation, and the remaining interactions for training. To ensure tractability and focus on recent behaviors, we set the maximum sequence length to 200. During testing, we rank the ground-truth item against all items, except in POI recommendation where only POIs within a $5$km radius of the user are considered. On this basis, we leverage two ranking metrics, namely Hit Ratio at Rank $k$ (HR@$k$) and Normalized Discounted Cumulative Gain at Rank $k$ (NDCG@$k$) at $k \in \{5, 10\}$ \cite{2007CoFiRank}. HR@$k$ measures whether the ground truth item appears in the top-$k$ results, while NDCG@$k$ further accounts for the position of the ground truth, assigning higher scores to higher-ranked correct predictions.

In our framework, we adopt LLaMA-3.1-8B as the cloud-side LLM and LLaMA-3.2-1B \cite{touvron2023llama} as the device-side SLM to support cloud-device collaborative recommendation. For fair comparison, all baseline methods that require an LLM backbone are implemented using LLaMA-3-8B to ensure consistency in model capacity. For hyperparameters, we set the dimension size to $1024$, the learning rate to $0.0002$, the dropout to $0.2$, the batch size to $64$, and the maximum training epoch to $20$. 
% think about the dimension size

\begin{table}[t]
\centering
\caption{Dataset Statistics}
\resizebox{\linewidth}{!}{
\begin{tabular}{lcccc}
\toprule
\textbf{Domain} & \textbf{Movie} & \textbf{Music} & \textbf{Sport} & \textbf{POI} \\
\textbf{Dataset} & ReDial & Music4All & Amazon & Foursquare \\
\midrule
\#User & 1,482 & 14,127 & 25,598 & 1,083 \\
\#Item & 33,834 & 109,269 & 18,357 & 38,333 \\
\#Interactions & 300,401 & 5,109,592 & 296,337 & 227,428 \\
\#Interactions/user & 202.70 & 361.69 & 11.58 & 210.00 \\
\#Tags & 37 & 57 & 42 & 38 \\
\bottomrule
\end{tabular}
}
\label{tab:ds}
\end{table}

% 可压缩 - 之后再决定
\subsection{Baselines}
We compare CDA4Rec against three types of recommendation frameworks, categorized based on their inference paradigms.

\noindent\textbf{Cloud-side Inference:}
\begin{itemize}[leftmargin=*]
  \item \textbf{GRU4Rec} \cite{hidasi2015session}: It is a classic sequential recommendation model that leverages gated recurrent units (GRUs) to capture user behavior patterns over time.
  
  \item \textbf{SASRec} \cite{kang2018self}: It is a self-attention based sequential recommendation model that adaptively captures both short-term and long-term user preferences by selectively attending to relevant past interactions.
  
  \item \textbf{LightGCN} \cite{he2020lightgcn}:It is a simplified graph-based collaborative filtering model that removes feature transformation and nonlinear activation, focusing solely on neighborhood aggregation to learn user and item embeddings with improved efficiency and performance.
  
  \item \textbf{SURGE} \cite{chang2021sequential}: It is a graph-based sequential recommendation framework that reconstructs user behavior sequences into interest-aware item graphs and performs cluster-aware and query-aware graph reasoning to dynamically extract users’ core preferences from noisy and long interaction histories.
  
    \item \textbf{RARec} \cite{yu2024ra}: It is a novel LLM-based recommendation framework that introduces an ID representation paradigm by aligning pre-trained ID embeddings with LLMs through soft prompts, achieving superior performance with minimal training data. In addition to the original cloud-based RARec (using LLaMA-8B), we also implement a lightweight version using LLaMA-1B as the on-device SLM baseline, denoted as \textbf{RARec*}.
    
    \item \textbf{TransRec} \cite{lin2024bridging}: It is a novel LLM-based recommendation paradigm that bridges items and language by using multi-facet identifiers (ID, title, attributes) and a structured grounding mechanism to enable accurate, semantic, and distinct item generation and ranking. Similarly, we include a lightweight variant with LLaMA-1B deployed on-device, denoted as \textbf{TransRec*}.

\end{itemize}

\noindent\textbf{Device-side Inference:}
\begin{itemize}[leftmargin=*]
  \item \textbf{LLRec} \cite{wang2020next}: It is a lightweight sequential recommender deployed on mobile devices, distilled from a powerful cloud-side teacher model.
  
  \item \textbf{PREFER} \cite{guo2021prefer}: It is a privacy-preserving and efficient federated framework for sequential recommendation that offloads training to user devices and aggregates model updates on edge servers.

  \item \textbf{DCCL} \cite{yao2021device}: It is a device-cloud collaborative learning framework that personalizes on-device models via MetaPatch and updates the centralized cloud model through a novel model-over-models distillation, enabling mutual enhancement across cloud and edge with strong performance on long-tailed users.
\end{itemize}

\noindent\textbf{Cloud-Device Collaborative Inference:}
\begin{itemize}[leftmargin=*]
  
  \item \textbf{DUET} \cite{lv2023duet}: It is a cloud-device collaborative framework for device model generalization that eliminates on-device fine-tuning by generating device-specific parameters through forward propagation, enabling efficient and accurate adaptation without incurring high computational costs.

  \item \textbf{LSC4Rec} \cite{lv2025collaboration}: It is a device-cloud collaborative recommendation framework that combines LLMs for semantic reasoning and SRMs for real-time adaptation, enabling efficient training and inference through collaborative training, inference, and intelligent request strategies.
\end{itemize}

\subsection{Overall Performance}

\begin{table*}[t]
\centering
\caption{Performance comparison across datasets and methods}
\resizebox{\textwidth}{!}{
\begin{tabular}{l|c@{\hskip 2pt}c@{\hskip 2pt}c@{\hskip 2pt}c@{\hskip 2pt}|c@{\hskip 2pt}c@{\hskip 2pt}c@{\hskip 2pt}c@{\hskip 2pt}|c@{\hskip 2pt}c@{\hskip 2pt}c@{\hskip 2pt}c@{\hskip 2pt}|c@{\hskip 2pt}c@{\hskip 2pt}c@{\hskip 2pt}c@{\hskip 2pt}}
\toprule
\textbf{Method} & \multicolumn{4}{c}{\textbf{Movie}} & \multicolumn{4}{c}{\textbf{Music}} & \multicolumn{4}{c}{\textbf{Sports}} & \multicolumn{4}{c}{\textbf{POI}} \\
\midrule
& {\fontsize{7}{9}\selectfont HR@5} & {\fontsize{7}{9}\selectfont NDCG@5} 
& {\fontsize{7}{9}\selectfont HR@10} & {\fontsize{7}{9}\selectfont NDCG@10} 
& {\fontsize{7}{9}\selectfont HR@5} & {\fontsize{7}{9}\selectfont NDCG@5} 
& {\fontsize{7}{9}\selectfont HR@10} & {\fontsize{7}{9}\selectfont NDCG@10} 
& {\fontsize{7}{9}\selectfont HR@5} & {\fontsize{7}{9}\selectfont NDCG@5} 
& {\fontsize{7}{9}\selectfont HR@10} & {\fontsize{7}{9}\selectfont NDCG@10} 
& {\fontsize{7}{9}\selectfont HR@5} & {\fontsize{7}{9}\selectfont NDCG@5} 
& {\fontsize{7}{9}\selectfont HR@10} & {\fontsize{7}{9}\selectfont NDCG@10} \\

\midrule
\multicolumn{17}{c}{\textbf{Cloud-side Inference}} \\
\midrule
GRU4Rec & 0.0122 & 0.0073 & 0.0232 & 0.0112 & 0.0105 & 0.0061 & 0.0185 & 0.0086 & 0.0089 & 0.0053 & 0.0162 & 0.0077 & 0.2571 & 0.1936 & 0.3678 & 0.2216 \\
SASRec & 0.0129 & 0.0077 & 0.0247 & 0.0119 & 0.0111 & 0.0064 & 0.0194 & 0.0091 & 0.0094 & 0.0057 & 0.0173 & 0.0081 & 0.2718 & 0.2070 & 0.3883 & 0.2329 \\
LightGCN & 0.0111 & 0.0069 & 0.0199 & 0.0099 & 0.0106 & 0.0062 & 0.0187 & 0.0088 & 0.0090 & 0.0053 & 0.0167 & 0.0079 & 0.2536 & 0.1876 & 0.3615 & 0.2177 \\
SURGE & 0.0118 & 0.0073 & 0.0212 & 0.0105 & 0.0113 & 0.0066 & 0.0199 & 0.0094 & 0.0095 & 0.0056 & 0.0177 & 0.0084 & 0.2683 & 0.1977 & 0.3836 & 0.2315 \\
RARec     & 0.0235 & 0.0166 & 0.0394 & 0.0240 & 0.0212 & 0.0142 & 0.0351 & 0.0206 & 0.0128 & 0.0090 & 0.0230 & 0.0125 & 0.3220 & 0.2328 & 0.4489 & 0.2669 \\
TransRec  & 0.0235 & 0.0171 & 0.0403 & 0.0247 & 0.0211 & 0.0146 & 0.0354 & 0.0212 & 0.0128 & 0.0091 & 0.0231 & 0.0129 & 0.3265 & 0.2396 & 0.4542 & 0.2735 \\
\midrule
\multicolumn{17}{c}{\textbf{Device-side Inference}} \\
\midrule
LLRec     & 0.0099 & 0.0059 & 0.0177 & 0.0085 & 0.0083 & 0.0048 & 0.0146 & 0.0069 & 0.0079 & 0.0046 & 0.0145 & 0.0066 & 0.2216 & 0.1613 & 0.3073 & 0.1950 \\
PREFER    & 0.0103 & 0.0065 & 0.0183 & 0.0092 & 0.0090 & 0.0053 & 0.0157 & 0.0074 & 0.0083 & 0.0050 & 0.0151 & 0.0070 & 0.2392 & 0.1768 & 0.3392 & 0.1974 \\
DCCL      & 0.0129 & 0.0075 & 0.0233 & 0.0119 & 0.0108 & 0.0066 & 0.0191 & 0.0091 & 0.0090 & 0.0059 & 0.0173 & 0.0086 & 0.2315 & 0.1751 & 0.3295 & 0.2112 \\
RARec*      & 0.0141	&0.0081	&0.0247	&0.0126	&0.0118	&0.0070	&0.0204	&0.0102	&0.0103	&0.0062	&0.0195	&0.0089	&0.2465	&0.1833	&0.3506	&0.2172 \\
TransRec*  & 0.0149	&0.0088	&0.0257	&0.0136	&0.0129	&0.0073	&0.0213	&0.0106	&0.0107	&0.0068	&0.0200	&0.0095	&0.2667	&0.1936	&0.3672	&0.2289 \\
\midrule
\multicolumn{17}{c}{\textbf{Cloud-device Collaborative Inference}} \\
\midrule
DUET      & 0.0172 & 0.0107 & 0.0285 & 0.0149 & 0.0135 & 0.0080 & 0.0227 & 0.0105 & 0.0109 & 0.0069 & 0.0195 & 0.0102 & 0.2872 & 0.2094 & 0.3908 & 0.2422 \\
LSC4Rec   & 0.0242 & 0.0170 & 0.0401 & 0.0253 & 0.0218 & 0.0150 & 0.0364 & 0.0214 & 0.0133 & 0.0094 & 0.0236 & 0.0130 & 0.3290 & 0.2387 & 0.4607 & 0.2799 \\
CDA4Rec     & 0.0259 & 0.0182 & 0.0423 & 0.0271 & 0.0245 & 0.0171 & 0.0398 & 0.0246 & 0.0197 & 0.0136 & 0.0304 & 0.0192 & 0.3589 & 0.2576 & 0.4845 & 0.3046 \\
\bottomrule
\end{tabular}
}
\label{tab:RQ1_accuracy}
\end{table*}

\subsubsection{Recommendation Accuracy}
Table \ref{tab:RQ1_accuracy} summarizes the recommendation accuracy of all sequential recommenders, where CDA4Rec consistently achieves the best performance. This clearly demonstrates the superiority of our collaborative framework, which dynamically allocates tasks between cloud and device based on user context. CDA4Rec not only leverages the powerful semantic reasoning of LLMs but also adapts to real-time device-side signals via structured modeling, achieving precise and timely personalization.

Among cloud-side methods, RNN-based models (e.g., GRU4Rec, SASRec) and graph-based models (e.g., LightGCN, SURGE) demonstrate complementary strengths across different domains. Graph-based models generally perform better in domains like music and sports, where item-item relationships are more informative, while RNN-based models are more effective in domains with stronger sequential patterns such as movies or POIs. Furthermore, the introduction of attention mechanisms gives SASRec and SURGE an additional advantage, allowing them to better model long-range dependencies and capture nuanced user intents, resulting in more robust overall performance.

RARec and TransRec achieve even higher accuracy than the above models by aligning item semantics with pretrained language models, enabling more expressive and generalizable user preference modeling. However, their dependence on large-scale LLMs raises concerns over inference latency and scalability, which can hinder real-time deployment. In contrast, on-device models such as LLRec, PREFER, and DCCL exhibit noticeable performance degradation. This is largely due to constrained model capacity and limited computing resources on edge devices, which hinder their ability to capture complex user patterns and deliver accurate predictions. To further examine this trade-off, we implement lightweight variants of RARec and TransRec using LLaMA-1B as the on-device SLM, denoted as \textbf{RARec*} and \textbf{TransRec*}. Despite retaining their original architectural designs, these variants show a substantial drop in performance, indicating that simply downsizing LLM-based methods is insufficient for achieving high-quality recommendation under strict device constraints.

Collaborative frameworks provide a middle ground, balancing accuracy and efficiency. DUET achieves fast adaptation by forwarding cloud-side parameters, but lacks fine-grained personalization. LSC4Rec, on the other hand, employs both LLMs and structured models across device and cloud, outperforming most baselines. However, its static pipeline and absence of adaptive user strategy planning limit its further gains. CDA4Rec outperforms LSC4Rec by introducing a context-aware strategy planner that tailors the collaboration path for each user scenario side inference for high-frequency users, thus realizing fine-grained personalization and efficient coordination.

Furthermore, we observe notable cross-domain patterns. The POI domain exhibits relatively high accuracy across all models, likely due to inherent geographical constraints that narrow the recommendation space. In contrast, the Sport domain shows lower overall accuracy due to extreme sparsity and limited interaction history. Notably, CDA4Rec maintains strong performance even in this challenging setting, demonstrating its strength in handling sparse data through adaptive strategy planning and fine-grained user modeling.

\subsubsection{Recommendation Efficiency}
Inference efficiency is crucial for real-world deployment, where low latency is essential for user experience and system scalability. As shown in Table \ref{tab:RQ1_efficiency}, We compare CDA4Rec with strong-performing baselines including RARec, TransRec and LSC4Rec. These models are selected for their competitive accuracy, while models with poor accuracy are excluded since they are not viable options in practical applications, regardless of their inference speed.

The results show that LLM-based methods such as RARec and TransRec incur high inference latency due to autoregressive decoding and complex token grounding, particularly in large-scale domains like Music. LSC4Rec, though leveraging cloud-device collaboration, still suffers from considerable overhead caused by its static pipeline and limited device-side optimization. In contrast, CDA4Rec achieves notably lower latency (i.e., under 100 ms per sample across all domains) by offloading lightweight tasks to the device and dynamically adjusting inference paths through a context-aware planner. This design minimizes redundant computation and communication, delivering efficient recommendations without compromising accuracy, and making CDA4Rec well-suited for real-world, latency-sensitive applications.

\begin{table}[ht]
\centering
\caption{Inference Time Comparison (ms/sample)}
\begin{tabular}{lcccc}
\toprule
\textbf{Model} & \textbf{Movie} & \textbf{Music} & \textbf{Sports} & \textbf{POI} \\
\midrule
RARec      & 493.18  & 1072.94 & 392.73 & 76.11 \\
TransRec   & 501.91  & 1115.98 & 411.13 & 79.24 \\
LSC4Rec    & 677.16  & 1500.72 & 531.32 & 82.35 \\
CDA4Rec      & \textbf{88.95}   & \textbf{93.41}   & \textbf{70.48}  & \textbf{42.61} \\
\bottomrule
\end{tabular}
\label{tab:RQ1_efficiency}
\end{table}

\subsection{Ablation Study (RQ2)}
To better understand the contribution of each component within the CDA4Rec framework, we conduct a series of ablation studies. As the core of CDA4Rec's adaptive capability, strategy planning dynamically configures the recommendation pipeline based on user context. To assess its effectiveness, we remove it (w/o. SP) and generate multiple fixed-strategy variants. The settings for these variants are summarized in Table \ref{tab:versions}, where each variant applies a static setting to the four key components of the pipeline: Semantic User Modeling, Candidate Retrieval, Structured User Modeling, and Final Ranking. In addition, we ablate the abstraction generation module (w/o. AG) by directly feeding the raw user query and interaction history into the LLM for strategy planning, bypassing the lightweight abstraction generated by the device-side SLM. This allows us to evaluate whether the learned user abstraction offers valuable information compared with raw context. We also disable the abstract adapter (w/o. AA) to assess its impact on abstract generation. Finally, we include LSC4Rec as a lightweight baseline for comparison. The results are shown in Table \ref{tab:RQ2}, which records the recommendation accuracy (HR@10) and inference time (millisecond per sample).

\begin{table}[ht]
\centering
\small
\caption{Ablation Settings on CDA4Rec Components}
\resizebox{\linewidth}{!}{
\begin{tabular}{lcccc}
\toprule
\textbf{Version} & \makecell[c]{Semantic\\User Modeling} & \makecell[c]{Candidate\\Retrieval} & \makecell[c]{Structured\\User Modeling} & \makecell[c]{Final\\Ranking} \\
\midrule
$v_1$  & $\alpha = 0.5$ & \{5000, 20\} & SASRec    & $\beta = 0.5$ \\
$v_2$  & $\alpha = 1$   & \{5000, 20\} & SASRec    & $\beta = 0.5$ \\
$v_3$  & $\alpha = 0$   & \{5000, 20\} & SASRec    & $\beta = 0.5$ \\
$v_4$  & $\alpha = 0.5$ & \{500, 3\}   & SASRec    & $\beta = 0.5$ \\
$v_5$  & $\alpha = 0.5$ & \{5000, 20\} & SURGE   & $\beta = 0.5$ \\
$v_6$  & $\alpha = 0.5$ & \{5000, 20\} & SASRec    & $\beta = 1$   \\
$v_7$  & $\alpha = 0.5$ & \{5000, 20\} & SASRec    & $\beta = 0$   \\
\bottomrule
\end{tabular}
}
\label{tab:versions}
\end{table}

\subsubsection{Effect of Strategy Planning}
Comparing $v_1$ ($\alpha=0.5$), $v_2$ ($\alpha=1$), and $v_3$ ($\alpha=0$) shows that balancing semantic and structured signals ($\alpha=0.5$) yields more stable performance across domains. Using only semantic information ($\alpha=1$) degrades accuracy, particularly in sparse domains like Music and POI. Conversely, discarding semantic signals ($\alpha=0$) also leads to performance drops, highlighting the complementarity of the two sources.

Compared to $v_1$, the variant $v_4$ uses a smaller candidate pool, achieving the lowest latency but also the worst accuracy on all datasets except POI. This demonstrates a trade-off where aggressive filtering reduces computational cost but harms recommendation quality. The exception in the POI dataset is due to its geographical constraint: candidates are already limited to within 5km of the user, making further reduction less detrimental.

A comparison between $v_1$ and $v_5$ indicates that replacing SASRec with SURGE brings a slight improvement on Music and Sport, but leads to minor declines on Movie and POI. This suggests that the two models exhibit different strengths across domains: SASRec is more effective in domains with clear sequential structures and stable user preferences, such as movies and location-based POIs, while SURGE performs better in dynamic domains like music and sports, where user interests shift frequently and item-item relational patterns play a larger role.

We further assess the effect of the final ranking stage by varying $\beta$. Setting $\beta=1$ in $v_6$ relies solely on the semantic score, while $\beta=0$ in $v_7$ depends only on structured behavior. $v_1$, which uses a hybrid strategy ($\beta=0.5$), consistently outperforms both, confirming the benefit of integrating heterogeneous signals.

Overall, these results indicate that different users and domains require different strategies at each stage of the pipeline to achieve optimal accuracy and efficiency, underscoring the critical importance of the proposed strategy planning module. This is further supported by the fact that CDA4Rec consistently outperforms LSC4Rec across all datasets.

\subsubsection{Effect of Abstract Generation}
To protect user privacy, CDA4Rec employs an on-device SLM to generate abstract representations of user intent, rather than exposing raw queries and behavioral histories to the cloud. While this abstraction introduces a slight drop in recommendation accuracy, the performance degradation is minimal and well within an acceptable range. Given the strong privacy benefits brought by this design, the trade-off is both reasonable and worthwhile.

To further enhance the quality of the generated abstractions, we incorporate a QLoRA-trained adapter into the SLM. This lightweight fine-tuning step effectively aligns the abstract outputs with the expectations of the cloud-side planner, helping to close the performance gap caused by abstraction. As a result, CDA4Rec is able to achieve strong accuracy while maintaining a privacy-preserving workflow.

\begin{table}[t]
\centering
\caption{Ablation study results (HR@10 and inference time in ms/sample) across datasets.}
\resizebox{\linewidth}{!}{
\begin{tabular}{l c@{\hskip 2pt}c@{\hskip 2pt}c@{\hskip 2pt}c@{\hskip 2pt} c@{\hskip 2pt}c@{\hskip 2pt}c@{\hskip 2pt}c@{\hskip 2pt} c@{\hskip 2pt}c@{\hskip 2pt}c@{\hskip 2pt}c@{\hskip 2pt} c@{\hskip 2pt}c@{\hskip 2pt}c@{\hskip 2pt}c@{\hskip 2pt}}
\toprule
\textbf{Method} & \multicolumn{2}{c}{\textbf{Movie}} & \multicolumn{2}{c}{\textbf{Music}} & \multicolumn{2}{c}{\textbf{Sport}} & \multicolumn{2}{c}{\textbf{POI}} \\
\cmidrule(lr){2-3} \cmidrule(lr){4-5} \cmidrule(lr){6-7} \cmidrule(lr){8-9}
 & HR@10 & ms & HR@10 & ms & HR@10 & ms & HR@10 & ms \\
\midrule
\textbf{CDA4Rec} & 0.0423 & 88.95 & 0.0398 & 93.41 & 0.0304 & 70.48 & 0.4845 & 82.35 \\
\textbf{LSC4Rec} & 0.0401 & 677.16 & 0.0364 & 1500.72 & 0.0236 & 531.32 & 0.4607 & 325.05 \\
\textbf{w/o AG} & 0.0415 & 76.12 & 0.0389 & 79.98 & 0.0295 & 62.24 & 0.4757 & 71.66 \\
\textbf{w/o AA} & 0.0380 & 87.44 & 0.0355 & 91.83 & 0.0271 & 69.22 & 0.4390 & 81.37 \\
\midrule
\multicolumn{9}{l}{\textbf{w/o SP}} \\
$v_1$ & 0.0415 & 89.16 & 0.0379 & 95.51 & 0.0239 & 70.76 & 0.4597 & 44.29 \\
$v_2$ & 0.0409 & 91.03 & 0.0337 & 94.89 & 0.0221 & 71.96 & 0.4471 & 44.01 \\
$v_3$ & 0.0358 & 91.17 & 0.0362 & 93.80 & 0.0235 & 71.07 & 0.4512 & 44.08 \\
$v_4$ & 0.0338 & 28.82 & 0.0281 & 31.67 & 0.0207 & 24.30 & 0.4517 & 14.27 \\
$v_5$ & 0.0412 & 91.81 & 0.0385 & 95.93 & 0.0243 & 72.23 & 0.4482 & 42.96 \\
$v_6$ & 0.0393 & 89.33 & 0.0315 & 97.06 & 0.0211 & 73.06 & 0.4172 & 42.93 \\
$v_7$ & 0.0377 & 89.76 & 0.0361 & 95.06 & 0.0232 & 71.16 & 0.4511 & 43.06 \\
\bottomrule
\end{tabular}
}
\label{tab:RQ2}
\end{table}

\subsection{Effect of Different Interaction History Lengths (RQ3)} % Done
\begin{figure}
    \centering
	\includegraphics[width=\linewidth]{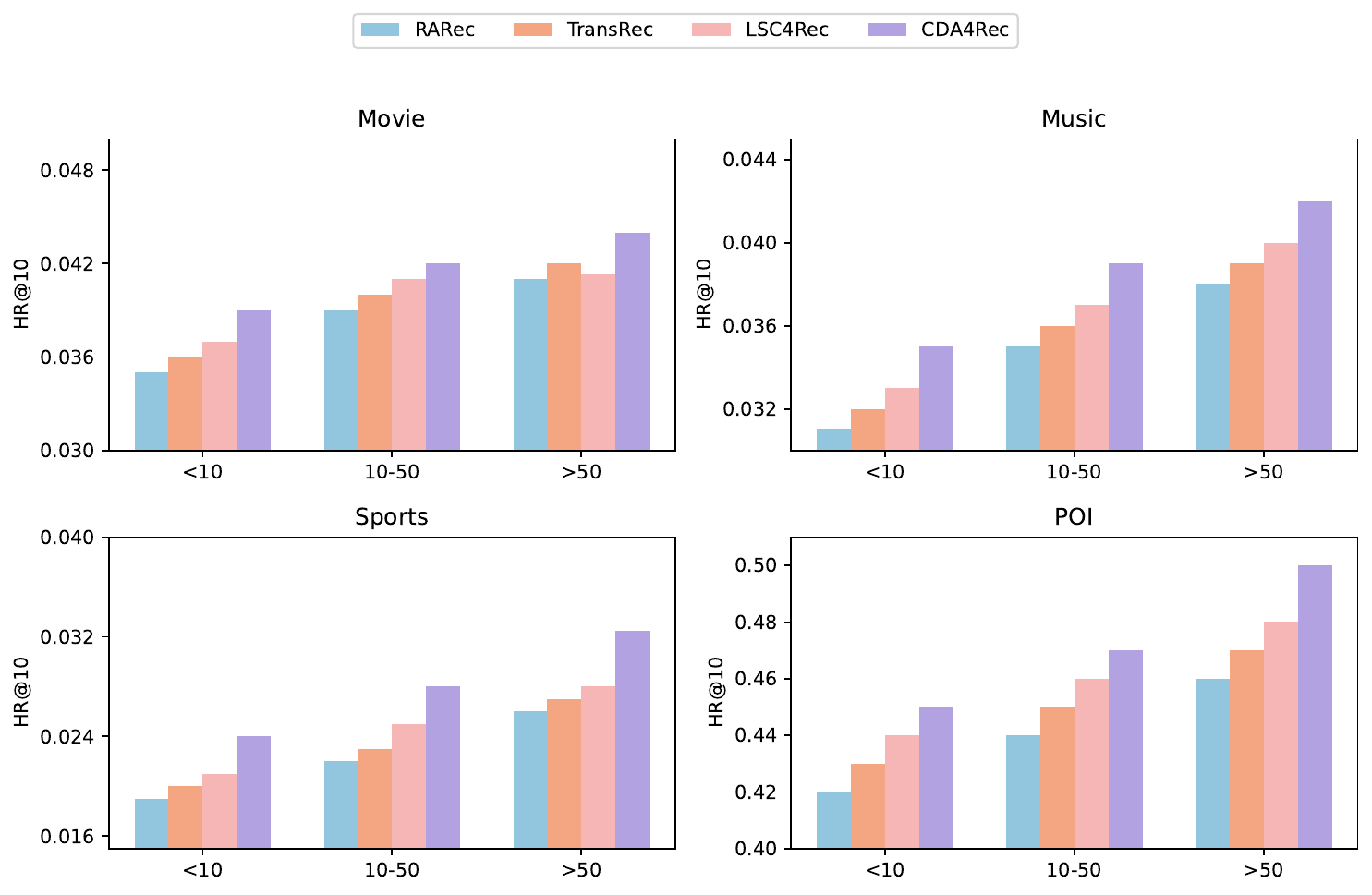}
	\caption{Recommendation performance w.r.t different interaction sizes.}
	\label{Fig:RQ3} 
\end{figure}

The length of a user user's interaction history significantly impacts recommendation accuracy, making it crucial for a recommender system to maintain robustness across varying history lengths. To systematically assess this effect, we categorize user's history lengths into three groups: short (fewer than 10 interactions), medium (10–50 interactions), and long (more than 50 interactions). We compare our method with three advanced baselines including RARec, TransRec, and LSC4Rec. The results are presented in Figure \ref{Fig:RQ3}, using HR@10 as the evaluation metric.

It is observed that recommendation accuracy improves consistently with longer interaction histories across all models, which aligns with the intuition that more behavioral signals facilitate better personalization. However, the performance gap between CDA4Rec and the baselines is particularly prominent in the short history group, which indicates CDA4Rec's strong ability to make accurate recommendations even with very limited user data. In the medium and long  history groups, CDA4Rec continues to outperform RARec, TransRec, and LSC4Rec, demonstrating its robustness and adaptability across different levels of user activity. Notably, while other models rely heavily on user-item co-occurrence patterns learned from sufficient history, CDA4Rec leverages its strategy planning and semantic modeling components to generalize effectively from limited signals.

\subsection{Few-shot Recommendation for Unseen Users (RQ4)}

\begin{table*}[t]
\centering
\caption{Few-shot recommendation performance for unseen users}
\resizebox{\textwidth}{!}{
\begin{tabular}{l|c@{\hskip 2pt}c@{\hskip 2pt}c@{\hskip 2pt}c@{\hskip 2pt}|c@{\hskip 2pt}c@{\hskip 2pt}c@{\hskip 2pt}c@{\hskip 2pt}|c@{\hskip 2pt}c@{\hskip 2pt}c@{\hskip 2pt}c@{\hskip 2pt}|c@{\hskip 2pt}c@{\hskip 2pt}c@{\hskip 2pt}c@{\hskip 2pt}}
\toprule
\textbf{Method} & \multicolumn{4}{c}{\textbf{Movie}} & \multicolumn{4}{c}{\textbf{Music}} & \multicolumn{4}{c}{\textbf{Sports}} & \multicolumn{4}{c}{\textbf{POI}} \\
\midrule
& {\fontsize{7}{9}\selectfont HR@5} & {\fontsize{7}{9}\selectfont NDCG@5} 
& {\fontsize{7}{9}\selectfont HR@10} & {\fontsize{7}{9}\selectfont NDCG@10} 
& {\fontsize{7}{9}\selectfont HR@5} & {\fontsize{7}{9}\selectfont NDCG@5} 
& {\fontsize{7}{9}\selectfont HR@10} & {\fontsize{7}{9}\selectfont NDCG@10} 
& {\fontsize{7}{9}\selectfont HR@5} & {\fontsize{7}{9}\selectfont NDCG@5} 
& {\fontsize{7}{9}\selectfont HR@10} & {\fontsize{7}{9}\selectfont NDCG@10} 
& {\fontsize{7}{9}\selectfont HR@5} & {\fontsize{7}{9}\selectfont NDCG@5} 
& {\fontsize{7}{9}\selectfont HR@10} & {\fontsize{7}{9}\selectfont NDCG@10} \\
\midrule
RARec     & 0.0207 & 0.0146 & 0.0347 & 0.0215 & 0.0208 & 0.0138 & 0.0349 & 0.0206 & 0.0113 & 0.0080 & 0.0203 & 0.0111 & 0.2865 & 0.2070 & 0.3959 & 0.2396 \\
TransRec  & 0.0207 & 0.0152 & 0.0357 & 0.0221 & 0.0207 & 0.0142 & 0.0349 & 0.0208 & 0.0115 & 0.0080 & 0.0204 & 0.0115 & 0.2910 & 0.2126 & 0.4086 & 0.2458 \\
LSC4Rec   & 0.0194 & 0.0139 & 0.0328 & 0.0203 & 0.0194 & 0.0131 & 0.0322 & 0.0191 & 0.0109 & 0.0075 & 0.0192 & 0.0104 & 0.2685 & 0.1943 & 0.3773 & 0.2267 \\
CDA4Rec     & \textbf{0.0248} & \textbf{0.0171} & \textbf{0.0404} & \textbf{0.0256} & \textbf{0.0241} & \textbf{0.0170} & \textbf{0.0408} & \textbf{0.0239} & \textbf{0.0186} & \textbf{0.0129} & \textbf{0.0286} & \textbf{0.0182} & \textbf{0.3417} & \textbf{0.2453} & \textbf{0.4610} & \textbf{0.2913} \\
\bottomrule
\end{tabular}
}
\label{tab:RQ4}
\end{table*}

In many practical applications, recommendation systems often encounter new users whose data has never been seen during training. This settings poses a critical challenge for generalization. To evaluate CDA4Rec’s generalization ability to such unseen users under limited behavioral evidence, we simulate a few-shot recommendation scenario. Specifically, we randomly hold out 10\% of users from each dataset, ensuring that none of their interaction records appear during training. During testing, we provide a short history of interactions and evaluate each model’s recommendation performance. We compare CDA4Rec with three advanced baselines: RARec, TransRec, and LSC4Rec, all of which are capable of handling few-shot or zero-shot scenarios to varying degrees.

As shown in Table \ref{tab:RQ4}, CDA4Rec consistently outperforms all compared baselines across datasets and evaluation metrics in the few-shot user setting. These improvements can be attributed to its strategy planning mechanism and semantic user modeling, which together allow the system to effectively infer user intent even from limited interaction data. While RARec and TransRec benefit from their recurrent modeling structures, and LSC4Rec performs reasonably well through lightweight abstraction, none of these approaches achieve the same level of consistency and accuracy as CDA4Rec. The performance gap is especially notable in sparse domain (i.e., sport), where interaction signals are limited and semantic generalization becomes crucial. These findings highlight CDA4Rec’s robust few-shot personalization capability, suggesting its strong potential for real-world recommendation scenarios where systems must frequently serve users whose data has not been seen during training.

\subsection{Case Study (RQ5)}
\begin{figure*}
	\includegraphics[width=\linewidth]{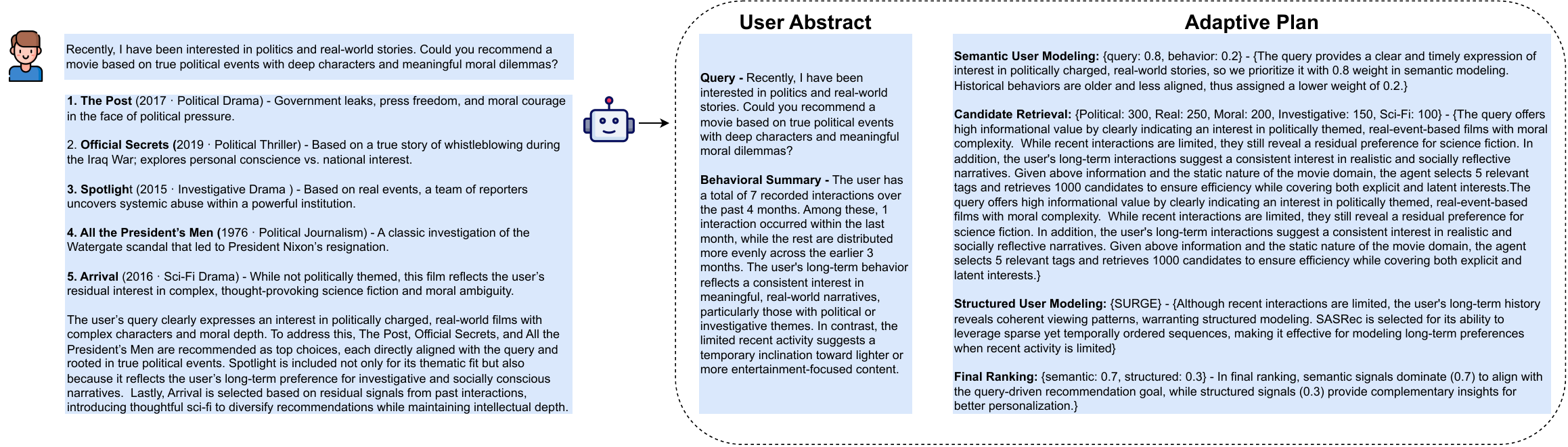}
    \caption{An example of Movie Recommendation.}
	\label{fig:case study} %% label for entire figure 
\end{figure*}

To further demonstrate the interpretability and adaptability of our CDA4Rec framework, we present a case study involving a real user query and interaction history. This example showcases how the device-side agent generates a privacy-preserving abstract, how the cloud agent plans a personalized recommendation strategy based on both the query and long-term behavioral patterns, and how the final recommendations reflect both the user's explicit intent and implicit preferences. The case illustrates CDA4Rec’s ability to dynamically tailor its recommendation pipeline while maintaining transparency throughout the decision-making process.

\section{Related Work}
This section reviews recent literature on related areas including sequential recommendation, on-device recommendation, and cloud-device collaborative recommendation.

\subsection{Sequential Recommendation}
Sequential recommendation aims to model users’ evolving preferences from their interaction histories. Early methods based on matrix factorization~\cite{mehta2017review} and Markov chains~\cite{rendle2010factorizing} focused on long-term preferences and short-term transitions but struggled with complex dependencies. To better capture sequential patterns, RNN-based models~\cite{hidasi2015session,xu2021long,cui2018mv} were introduced to learn local temporal dynamics through hidden state updates. Graph-based methods~\cite{liu2024selfgnn,chang2021sequential,zhang2022dynamic} further modeled item transitions as session graphs, capturing high-order and structural dependencies. Attention mechanisms~\cite{kang2018self,li2020time} have since been adopted as flexible modules that enhance RNNs and GNNs, or serve as standalone architectures, with self-attention particularly effective in modeling global dependencies and enabling parallel, context-aware recommendation.

More recently, large language models (LLMs) have been applied to sequential recommendation due to their strong generative and reasoning capabilities. Some works reformulate recommendation as a natural language task, prompting pretrained LLMs with user histories and item metadata to enable zero-shot or few-shot prediction~\cite{lin2024data, geng2022recommendation, li2023prompt}. Additionally, ID-based LLM approaches \cite{yu2024ra,lin2024bridging} directly model sequences of item IDs or embeddings without relying on textual descriptions, offering better compatibility with traditional recommendation pipelines and improved efficiency in structured domains. Building on these foundations, recent studies \cite{shu2024rah, zhang2024generative} have begun exploring LLM-based agent frameworks for recommendation. These approaches decompose the recommendation process into modular sub-tasks and assign them to specialized LLM agents that collaborate through message passing or shared memory. Our work build this line of research and extend the LLM-agent paradigm to the cloud-device setting, aiming to enable flexible, personalized, and interpretable sequential recommendation under real-world constraints.

\subsection{On-Device Recommendation}
On-device recommendation performs the entire inference process locally on the user's device, offering the advantage of reduced latency without relying on cloud-based computation. To enable effective on-device recommendation under resource constraints, various techniques have been developed. One line of work focuses on model compression \cite{wang2020next, yao2021device}, including knowledge distillation, pruning, and quantization, to deploy compact versions of high-capacity recommendation models. Another approach leverages federated learning \cite{yang2020federated, qu2024towards}, where the model is collaboratively trained across multiple devices without transmitting raw user data, thus preserving privacy while capturing cross-user knowledge. More recent research \cite{long2023decentralized, long2023model} explores decentralized or peer-to-peer training paradigms, where devices share updates directly with each other, enabling personalized model adaptation through local communication. Despite these advances, on-device recommendation remains limited by computational and memory constraints, making it difficult to support complex architectures or large-scale retrieval, and often leading to suboptimal personalization in practice.

\subsection{Cloud-Device Collaborative Recommendation}
Cloud-device collaborative recommendation aims to combine the modeling power of cloud-based systems with the low-latency and privacy benefits of on-device inference. By splitting the recommendation pipeline, these frameworks enable more responsive and context-aware recommendation. For instance, MetaController \cite{yao2022device} introduced dynamic collaboration strategies that allow the system to switch between cloud and device recommenders based on user behavior patterns. DUET \cite{lv2023duet} and IntellectReq \cite{lv2024intelligent} focused on enhancing the generalization ability of device models by enabling the cloud to generate personalized parameters or determine communication needs using uncertainty estimation. Finally, LSC4Rec \cite{lv2025collaboration} proposed a collaborative framework that integrates a cloud-based LLM with an on-device SRM, supporting joint training and inference to leverage the strengths of both models across cloud and edge environments. However, most existing methods rely on fixed collaboration patterns and lack the flexibility to adapt to diverse user contexts. In contrast, our approach introduces a dynamic and interpretable framework that adjusts cloud-device cooperation based on user-specific signals.

\section{Conclusion}
In this paper, we introduced CDA4Rec, a cloud-device collaborative framework for sequential recommendation that addresses key challenges in personalization, efficiency, and privacy. Built upon a dual-agent architecture, CDA4Rec assigns semantic modeling and candidate retrieval to the cloud agent, while delegating structured modeling and final ranking to the device agent. A strategy planning mechanism orchestrates this collaboration by generating personalized execution plans based on user context. This design enables adaptive task division and partial parallelism across agents, significantly improving system responsiveness. Extensive experiments across multiple domains validate the superiority of CDA4Rec over strong baselines, particularly under diverse user states and real-world constraints.

\bibliographystyle{IEEEtran}
\bibliography{main}

\vfill
\end{document}